\documentclass[9pt,twocolumn,twoside]{pnas-new}

\templatetype{pnasresearcharticle} 

\title{High-throughput measurement of elastic moduli of microfibers by rope coiling}

\author[a,1]{Yuan Liu}
\author[b,1]{Jack H. Y. Lo}
\author[c]{Janine K. Nunes}
\author[c,2]{Howard A. Stone}
\author[a,2]{Ho Cheung Shum}

\affil[a]{Department of Mechanical Engineering, University of Hong Kong, Hong Kong SAR, China}
\affil[b]{Center for Integrative Petroleum Research, College of Petroleum Engineering and Geosciences, King Fahd University of Petroleum and Minerals, Dhahran, Saudi Arabia}
\affil[c]{Department of Mechanical and Aerospace Engineering, Princeton University, Princeton, New Jersey, USA}

\leadauthor{Liu\&Lo}

\significancestatement{Fibers are ubiquitous in nature. They are the building blocks of various advanced materials and natural structures, such as the cytoskeleton, artificial muscles, filamentous bacteria, DNA, carbon nanotubes, spider silk, optical fibers, etc. Their mechanical performance largely depends on the elastic modulus. Typically, the elastic modulus is measured by holding one end under a given load, pulling the other end, and observing the deformation. However, for tiny fragile samples, it is very slow to measure them in quantity, owing to the difficult and time-consuming sample loading/unloading. Here, we develop a simple microfluidic method that continuously measures the elastic modulus with high throughput using rope coiling. This is based on the quantitative correlation between the elastic modulus and the coiling radius.}

\authorcontributions{Y.L., J.H.Y.L., J.K.N., and H.A.S. conceived and designed the study; Y.L. and J.H.Y.L. performed the experiments; all authors analyzed the data; Y.L., J.H.Y.L., H.A.S., and H.C.S. wrote the manuscript.}
\authordeclaration{Y.L., H.C.S., J.K.N., and H.A.S. are co-inventors of a filed US patent no.63/367,173, which describes the methods used herein.}
\equalauthors{\textsuperscript{1}Y.L. and J.H.Y.L. contributed equally to this work.}
\correspondingauthor{\textsuperscript{2}To whom correspondence should be addressed. E-mail: hastone@princeton.edu;ashum@hku.hk}


\begin{abstract}
There are many fields where it is of interest to measure the elastic moduli of tiny fragile fibers, such as filamentous bacteria, actin filaments, DNA, carbon nanotubes, and functional microfibers. The elastic modulus is typically deduced from a sophisticated tensile test under a microscope, but the throughput is low and limited by the time-consuming and skill-intensive sample loading/unloading. Here we demonstrate a simple microfluidic method enabling the high-throughput measurement of the elastic moduli of microfibers by rope coiling using a localized compression, where sample loading/unloading are not needed between consecutive measurements. The rope coiling phenomenon occurs spontaneously when a microfiber flows from a small channel into a wide channel. The elastic modulus is determined by measuring either the buckling length or the coiling radius. The throughput of this method, currently 3300 fibers per hour, is a thousand times higher than that of a tensile tester. We demonstrate the feasibility of the method by testing a non-uniform fiber with axially varying elastic modulus. We also demonstrate its capability for in-situ inline measurement in a microfluidic production line. We envisage that high-throughput measurements may facilitate potential applications such as screening or sorting by mechanical properties and real-time control during production of microfibers.
\end{abstract}

\dates{This manuscript was compiled on \today}
\doi{\url{www.pnas.org/cgi/doi/10.1073/pnas.XXXXXXXXXX}}

\begin{document}

\maketitle
\thispagestyle{firststyle}
\ifthenelse{\boolean{shortarticle}}{\ifthenelse{\boolean{singlecolumn}}{\abscontentformatted}{\abscontent}}{}

\dropcap{F}ibers are ubiquitous in physical and biological systems; they are the building blocks of various advanced materials and many natural structures are in the form of fibers or fiber networks. Their elastic properties are of considerable interest: For cells or rod-shaped bacteria, elastic properties of the actin filament network or the cell wall influence functions such as cell adhesion, migration, and growth \cite{amir2014bending,gardel2004elastic,wu2018comparison,dufrene2020mechanomicrobiology}. For DNA, it helps understand the structural dynamics of cellular processes such as replication and transcription \cite{bustamante2003ten,camunas2016elastic}. For a hydrogel, as used in wound dressings, it is desirable to minimize stress induced on a wound \cite{perazzo2017flow,magnani2021recent,shen2023fibro}. Optical fibers with high flexibility and stretchability are useful for invasive, implantable, or wearable medical devices \cite{guo2016highly,sarabi2021biomedical}, while spider silk and carbon nanotubes are promising textile materials due to their outstanding mechanical properties \cite{wong1997nanobeam,yu2000strength,zhang2004multifunctional}, and, finally, fiber-shaped actuators are designed as artificial muscles that could propel biomimetic robots \cite{tawfick2019stronger}. It is clear that the materials and applications are diverse, and their elastic properties have a great impact.

The elastic modulus is commonly measured by a tensile test: fixing one end of a sample, pulling the other end under a given load, and measuring the sample’s deformation. Using similar ideas, the elastic moduli of tiny fragile fibers, such as filamentous bacteria \cite{amir2014bending,boulbitch2000elasticity}, actin filaments \cite{kojima1994direct}, DNA \cite{bustamante2003ten,shon2019submicrometer}, carbon nanotubes \cite{yu2000strength,poncharal1999electrostatic,demczyk2002direct}, and functional microfibers \cite{guo2016highly,duprat2015microfluidic,nunes2021electrostatic} can be measured by advanced equipment such as optical/magnetic tweezers \cite{bustamante2003ten,shon2019submicrometer}, piezo actuators \cite{guo2016highly,kojima1994direct,demczyk2002direct}, atomic force microscopes \cite{yu2000strength,alsteens2013multiparametric}, transmission electron microscopes \cite{poncharal1999electrostatic,demczyk2002direct}, and microfluidic devices \cite{amir2014bending,duprat2015microfluidic,nunes2021electrostatic}. Recent technological advances in these measurements prevail on its sensitivity and accuracy ($\sim$pN in force and $\sim$nm in displacement). The throughput, however, is low – it is impeded by the sample loading and unloading, which is time-consuming and skill-intensive as many of the fibers of interest are commonly small and fragile. In general, increasing the throughput not only reduces the time and labor costs, it may open the door to potential applications and bring substantial impact, as shown by the success in next-generation sequencing and high-throughput screening in biomedical research.

Here we demonstrate a simple microfluidic method that measures the elastic moduli of microfibers at high throughput by observing the coiling of the filament following a localized compression, where sample loading and unloading are not needed between consecutive measurements. We use the term "rope coiling" to refer to the post-buckling deformation of an elastic fiber \cite{du2019dynamics}. It should not be confused with the coiling of viscous liquid threads \cite{mahadevan1998fluid,maleki2004liquid,cubaud2006folding,ribe2012liquid,tian2020steady}, nor the subsequent solidification of the already coiled liquid threads \cite{xu2017bioinspired,shao2018fiber}. The interest of rope coiling in liquid has emerged only recently with increasing microfluidic applications. Notably, the rope coiling or buckling in liquid has been studied and demonstrated in simulations \cite{chelakkot2012flow} and in experiments on metal wire \cite{gosselin2014buckling}, actin filaments \cite{chakrabarti2020flexible}, PEG-based hydrogel fibers \cite{cappello2022fiber}, and polyvinyl alcohol-carbon nanotube composite fibers \cite{mercader2010kinetics}, which inform the current study. Nevertheless, the basic configuration of this study is different from some of the previous studies. In our work, fibers are carried by the surrounding flow and move continuously, while in the work of Gosselin et al. a wire is injected from air into a quiescent liquid bath \cite{gosselin2014buckling}; similarly, in the study of Cappello et al. both the fiber and the flow are initially at rest \cite{cappello2022fiber}. In our work, the compression is localized or temporary, while in the work of Chakrabarti et al. the compression is continuous and constant \cite{chakrabarti2020flexible}. The differences in configurations lead to different results.

\section*{Results}
\subsection*{1. Experimental setup and coiling of microfibers}

We first demonstrate that rope coiling occurs when a microfiber flows through a device, shown in Fig. 1\textit{A}, \textit{B}, which is made by connecting two glass capillaries to form a small channel upstream and an abruptly enlarged wide channel downstream. The small channel is a circular tube, while the wide channel is a square tube to minimize image distortion. The setup and the flow are approximately axisymmetric (see SI Appendix), the resulting coiling is three-dimensional rather than two-dimensional as found in a related study \cite{gosselin2014buckling}. The microfiber we use as a model is made of poly(ethylene glycol) diacrylate and prepared by a separate microfluidic device. The surrounding medium is water with 1 wt\% of tripotassium phosphate salt (see Materials and Methods).

Next, to control experimental conditions precisely, we attach the coiling device directly to a device for making fibers, which can produce microfibers on the fly with different diameters, elastic moduli, and injection speeds, as shown in Fig. 1\textit{C, D}; elastic moduli are controlled by the intensities of UV light (see Materials and Methods). The microfiber is straight before entering the wide channel in the coiling device as shown in Fig. 1\textit{E}. Separately, we measure the elastic moduli of microfibers by a tensile tester (see Methods and SI Appendix).

Microfibers coil spontaneously when they enter the wide channel downstream, and the resulting geometry of the filament is influenced by the elastic modulus, as shown in Movie S1. In our typical experiments, the injection speed of the fibers is 42 mm/s, the length of a fiber is 40 mm, and the time between consecutive measurements is 0.1 s. The corresponding throughout is 3300 fibers per hour or $\approx$0.9 fibers per second. The current device can run continuously for at least 90 min, limited by the injectable volume of the syringe pump. About 5000 microfibers are collected in a glass vial as shown in Fig. 1\textit{G}. The coiling geometry is reproducible, as demonstrated by the continuous injection of fibers with the same size and same elastic moduli as shown in Movie S2.

As the fiber elastic modulus increases (from left to right in Fig. 1\textit{F} and Movie S3), the coiling radius increases. There is no coiling of a liquid jet under the same condition(Fig. 1\textit{F}, first column, $E$ = 0), indicating that we are observing the phenomenon of elastic coiling \cite{chelakkot2012flow,gosselin2014buckling,chakrabarti2020flexible,cappello2022fiber,mercader2010kinetics,habibi2007coiling,jawed2014coiling} rather than viscous coiling \cite{mahadevan1998fluid,maleki2004liquid,cubaud2006folding,ribe2012liquid,tian2020steady}.

The coiling of fibers in the wide channel is temporary such that their natural curvature is zero. Because the fibers are very soft, sometimes they appear to curl due to boundary confinement. We have carefully suspended a fiber vertically in water, ensuring that it does not come into contact with the container or other fibers, as shown in Fig. 1\textit{H}. In the absence of friction, the fiber appears straight as expected. As the fiber is only 2\% denser than water, buoyancy has balanced most of its weight. This observation demonstrates that the microfibers have been fully solidified before coiling; otherwise, the elastic microfibers would have a curly shape at rest, as demonstrated in previous studies \cite{nunes2013microfluidic,chakrabarti2021instabilities}. This also justifies that the coiling does not induce plastic deformation (see SI Appendix for further information). In addition, we have shown that viscoelastic effects are negligible in our experiments, as detailed in the SI Appendix (Fig. S6).

\subsection*{2. General characteristics of the system}

We attribute the fiber coiling to the deceleration of the surrounding flow in the wide channel, which exerts a compressive stress on the fiber. We characterize the surrounding flow in the coiling device by experiments and calculations via COMSOL. The experimental flow visualization is shown in Fig. 2\textit{A} (see Materials and Methods). The numerically calculated flow velocity magnitude $|\vec{u}|$ and streamlines in the absence of fibers are shown in Fig. 2\textit{B}. The measured and calculated streamlines agree with each other and corner eddies are observed due to the sudden enlargement of the channel. The velocity data in Fig. 2\textit{B} along the central line (the axis of symmetry), denoted as $u_0\equiv|\vec{u}\left(r=0\right)|$, is plotted in Fig. 2\textit{E}. Near the entrance of the wide channel, the flow velocity decreases rapidly. The local Reynolds number of fiber is Re = $\frac{\rho v_{\mathrm{in}} d}{\mu}$, where $\rho$ is the density of the surrounding fluid, $v_{\mathrm{in}}$ is the injected velocity of fiber, $d$ is the fiber diameter, and $\mu$ is the viscosity of fluid. In our experiments, Re=O(1), ranges from 0.8 to 4.7, and it is much smaller than the slenderness ratio $k$$\sim$300. The slenderness ratio is defined as the fiber’s length-to-radius ratio. Therefore, the fiber experiences a large Stokes drag and compressive stress.

The rod-to-coil transition happens in this compressive region, indicated by the plot of the axial strain rate $\dot{\epsilon}\equiv d\vec{u}/dz\cdot\hat{z}$ in Fig. 2\textit{C} and the photo of a coiling microfiber in Fig. 2\textit{D}. The same data along the central line, ${\dot{\epsilon}}_0\left(z\right)\equiv du_0/dz$, is plotted in Fig. 2\textit{E}. Beyond the compressive region, where most of the helical coils are found, the radial velocity profile is parabolic as expected (see also SI Appendix Fig. S1-S3).

A timelapse photo of the coils is shown in the inset of Fig. 2\textit{E}, where the silhouette of the coils (red dotted lines) marks the evolution of the coiling radius $R_{\mathrm{coil}}$. While the strain rate increases and then decreases, the coiling radius increases steadily during compression. When the localized compression effectively vanishes, the coiling radius becomes constant.

Snapshots of the rod-to-coil transition are shown in Fig. 2\textit{F}, where the fiber first buckles and then winds into the first coil. We define $t$ = 0 as the moment when the first coil is formed. The subsequent motion beyond $t$ = 0 is periodic.

The coils of the microfibers formed at different times (rows) with different elastic moduli (columns) are shown in Fig. 2\textit{G}. All other conditions are the same. The coiling radii $R_{\mathrm{coil}}$ are marked by the red dotted lines. As the elastic modulus $E$ increases, the coiling radius $R_{\mathrm{coil}}$ increases. Moreover, the coiling radius remains constant with time and within a certain distance, as shown in Fig. 2\textit{H} and \textit{I}, so that it can be readily measured without the need for sophisticated instruments and techniques.

\subsection*{3. Measurement of elastic modulus}

The elastic moduli of the microfibers can be determined from either the buckling length or the coiling radius, with different advantages and disadvantages. The buckling of the rod-like fibers (see Fig. 2\textit{F} and inset of Fig. 3\textit{A}) can be explained by considering the bending rigidity and the viscous drag. The Euler critical load per length is estimated as\cite{hibbeler2018mechanics}:
\begin{equation}
f_c=\frac{\pi^2B}{{K^2\ L}^3}
\end{equation}
where $B = EI$ is the bending rigidity, $E$ is Young’s elastic modulus,  $I = \pi d^4/64$ is the area moment of inertia, $d$ is the fiber diameter, $L$ is the buckling length of the microfiber (i.e., the axial distance over which the fiber starts to buckle), and $K$ is an effective-length factor that depends on the boundary conditions. By measuring the distance between zero moment points, we infer that $K$$\sim$0.7, resembling the buckling with one fixed end and one pinned end (see SI Appendix Fig. S7). The viscous drag per length acting of the fiber is estimated as\cite{liu2018morphological,powers2010dynamics,batchelor1970slender}:
\begin{equation}
f_\mu=\frac{2\pi}{\ln{k}}\ \mu v_{\mathrm{in}}
\end{equation}
where $v_{\mathrm{in}}$ is the injected velocity of fiber, and $\mu$ is the viscosity of the surrounding fluid, adn $k$$\sim$300 is the fiber's slenderness ratio. Combining equations (1) with (2), the buckling length is deduced as:
\begin{equation}
L=\alpha\left(\frac{B}{\mu v_{\mathrm{in}}}\right)^{1/3}
\end{equation}
where $\alpha=\left(\frac{\pi}{2K^2}\ln{k}\right)^{1/3}$$\sim$2.6. A plot of $L$ versus
$\left(\frac{B}{\mu v_{\mathrm{in}}}\right)^{1/3}$ is shown in Fig. 3\textit{A} and exhibits a linear relation as expected. The slope of the best fit is $\alpha_{\mathrm{exp}}$ = 1.02, which is in agreement with but slightly less than the expected value above. The results align with prior studies in which the elastic modulus is held constant while the diameters or velocities vary\cite{gosselin2014buckling,chakrabarti2021instabilities}. Previous research has suggested that the coiling radius should follow the same scaling as\cite{chelakkot2012flow,chakrabarti2021instabilities}:
\begin{equation}
R_{\mathrm{coil}}=\beta\left(\frac{B}{\mu v_{\mathrm{in}}}\right)^{1/3}
\end{equation}

As verified in Fig. 3\textit{B}, the data are consistent with the theory with a best fit of $\beta$ = 0.33, though the data are more scattered when compared with that of the buckling length in Fig. 3\textit{A}, which will be addressed further below.

The elastic modulus can be obtained by measuring either the buckling length in (3) or the coiling radius in (4). The former approach appears more promising due to its theoretical robustness and its better alignment with experimental data in Fig. 3\textit{A}. However, it has a limitation in that the buckling length only reflects the local elastic modulus of the initial segment of the fiber, making it effective only for fibers with a uniform elastic modulus. In contrast, we will demonstrate that the local elastic modulus of a non-uniform fiber can be deduced from the coiling radius. Another advantage of using coiling radius instead of the buckling length is its ease of measurement and analysis because the coiling radius can remain constant over a certain distance and time (Fig. 2), but the buckling length can only be inferred in a short period of time.

Next, we will demonstrate the feasibility of measuring the coiling radius to determine the elastic modulus of a non-uniform fiber. We prepared a non-uniform fiber whose elastic modulus varies along its length, as illustrated in Fig. 3\textit{C}, by increasing the intensity of the UV light linearly with time during fiber production. The profile of the elastic modulus $E(s)$ is calculated and represented as a solid curve in Fig. 3\textit{D}, where $s$ is the arclength from the soft end (see SI Appendix Fig. S8 and reference therein\cite{duprat2015microfluidic}). As the non-uniform fiber coils in the microfluidic device, it forms a spiral with a varying coiling radius that reflects the local elastic modulus. The higher the local elastic modulus, the larger the coiling radius of that segment (Movie S4).

Using the buckling length and (3), the elastic modulus of the initial segment of the fiber tip is accurately determined (rhombus in Fig. 3\textit{D}), but this does not provide information about the rest of the fiber. Using the coiling radius and (4), the elastic modulus of the entire fiber is determined. However, the accuracy is unsatisfactory (squares in Fig. 3\textit{D}), aligning with the observed scatter in the data in Fig. 3\textit{B}. To improve the calibration, we introduce an extra phenomenological parameter such that (4) is revised as:
\begin{equation}
E=\gamma\frac{64}{\pi}\mu v_{\mathrm{in}}{R_{\mathrm{coil}}}^3/d^4+\delta
\end{equation}
where the slope $\gamma$ = 9.15$\pm$0.42 and intercept $\delta$ = 4.0$\pm$0.2 are determined by fitting of a set of calibration data, which have the same $d$ and $v_{\mathrm{in}}$ but different $E$ as the current test, as shown in the inset of Fig. 3\textit{D}. In other words, all fitting parameters are solely derived from the calibration data and independent of the measurement. The result is satisfactory (circles in Fig. 3\textit{D}), providing the elastic modulus of the entire fiber with high accuracy. The elastic moduli measured from $R_{\mathrm{coil}}$ are discrete because the camera image can only provide the coiling radius of the left-most and right-most points (see the rightmost side of Fig. 3\textit{C}). The distances of the data points are calculated iteratively by $s_{i+1}=s_{i}+\pi(R_{i+1}+R_{i})/2$ and $s_{1}=2\pi R_{1}$. The semi-empirical approach, using Eq. 5, has successfully proven its feasibility, where the accuracy of measurement is tied to the quality of the calibration experiments.

\subsection*{4. Geometric parameters}

The coiling system can be divided into three parts by geometry: the straight tail, the conical spiral, and the circular helix, as shown in the snapshot and time-lapse photo in Fig. 4\textit{A}. It is easy to confuse the spiral with the helix. Note that the existence of the helix is a signature of our system, and the helix is the primary focus of previous analysis. In contrast, when considering coiling on solid substrate, the primary focus is the spiral\cite{ribe2012liquid}. The coiling radius discussed in this paper refers to that of the circular helix, or equivalently, that of the end part of the spiral.

Elements on the spiral are rotating around the central axis but those on the helix do not. This is revealed by particle tracking experiments in which tracer particles are embedded in the fibers (see Fig. 4\textit{B} and Materials and Methods). For the helix, the tracers are simply translating downwards, leaving straight vertical trajectories. For the spiral, the tracers draw a spiral path along the surface of an imaginary cone.

The geometry of the helix, in general, is fully described by the coiling radius $R_{\mathrm{coil}}$, the frequency $\Omega$ (or equivalently the period $T$) of coiling, the downward translational speed $v$, and the pitch $P$, as shown in Fig. 4\textit{C}. We will elucidate the relationship between these parameters and investigate whether they can be used to deduce the elastic modulus, serving as an alternative for the coiling radius: Our case is slightly different from the coiling on a solid substrate, where $\Omega=\frac{v_{\mathrm{in}}}{R_{\mathrm{coil}}}$ and $v$ = $P$ =0, as the coils are forced to stop and stack on the solid substrate. Instead, in our case, $v$ is finite, $P=\frac{2\pi v}{\Omega}$ and $\Omega=\frac{2\pi v_{\mathrm{in}}}{\sqrt{P^2+\left(2\pi R_{\mathrm{coil}}\right)^2}}$. We hypothesize that the coils are moving at the same speed as the surrounding flow, such that $\bar{v}=\left(1-{\bar{R}}_{\mathrm{coil}}^2\right)$, which is the typical parabolic profile of Poiseuille flow in a tube, where $\bar{v}\equiv\frac{v}{u_{0}}$, ${\bar{R}}_{\mathrm{coil}}\equiv\frac{R_{\mathrm{coil}}}{R_{0}}$, $R_{0}$ is the half-width of the wide channel, and $u_{\mathrm{0}}$ is the flow velocity at the central axis. Accordingly, if $v\ll v_{\mathrm{in}}$, the expressions simplify to $\bar{P}\approx{\bar{R}}_{\mathrm{coil}}\left(1-{\bar{R}}_{\mathrm{coil}}^2\right)$ and $\ \Omega\approx\frac{v_{\mathrm{in}}}{R_{\mathrm{coil}}}$, where $\bar{P}\equiv\frac{P}{P_0}$ and $P_0\equiv2\pi R_0u_0/v_{\mathrm{in}}$. The expression of coiling frequency becomes identical to that of coiling on a solid substrate, provided that the channel is wide enough such that $v\ll v_{\mathrm{in}}$ or, consequently, $P\ll R_{\mathrm{coil}}$.

The experimental data of different conditions agree well with the expressions deduced above, as shown in Fig. 4\textit{D-F}: all data collapse on the same line.  There are no fitting parameters in the plots and $\frac{v}{v_{\mathrm{in}}}<0.1$ in our experiments. The coiling frequency in Fig. 4\textit{D} is plotted as $\bar{\Omega}\approx1/{\bar{R}}_{\mathrm{coil}}$, where $\bar{\Omega}\equiv\frac{\Omega}{\Omega_0}$ and $\Omega_0=\frac{v_{\mathrm{in}}}{R_0}$. The frequency $\Omega$ is obtained by measuring the coiling period $T$ for increasing the number of coils. The translational speed of coil $v$ in Fig. 4\textit{E} is measured from the distance traveled by a coil within a finite time (see SI Appendix Fig. S9). The values of $u_0$ are calculated from numerical simulations at $z$=1 mm, 1.5 mm from the entrance for $R_0$=580 \textmu m, 873 \textmu m respectively, where the flow velocity is almost constant (see Fig. 2\textit{B} and SI Appendix).

It is interesting that the pitch $\bar{P}$ does not increase monotonically with ${\bar{R}}_{\mathrm{coil}}$ (or $E$) but peaks at ${\bar{R}}_{\mathrm{coil}}=1/\sqrt3$. This implies that it is inappropriate to deduce $E$ from $\bar{P}$. This happens only in microfluidic channels as the flow is affected by the boundaries. For the case of a quiescent liquid bath, it has been shown that the pitch of the coils increases monotonically\cite{chakrabarti2021instabilities}.

The coiling frequency can be used to deduce $E$ in the same way as $R_{\mathrm{coil}}$ does. The advantage is that the coiling frequency is frequently used as the main variable in theoretical models\cite{ribe2012liquid,habibi2007coiling}. The disadvantage is that $R_{\mathrm{coil}}$ may be more straightforward to obtain than $\Omega$ in measurements, but in principle, $\Omega$ can also be directly measured.

\subsection*{5. Force analysis of the late stage of the helix}

The typical force balance between elastic force, viscous force, and centrifugal force\cite{habibi2007coiling,powers2010dynamics} is absent in the later stage of coiling. The magnitude of the elastic force per length is \(f_E\)$\sim$\(\frac{Ed^4}{{R_{\mathrm{coil}}}^3}\)$\sim$\({10}^{-2}\) Pa·m. In comparison, however, both the centrifugal force per length, \(f_I\)$\sim$\(\frac{\rho d^2U^2}{R_{\mathrm{coil}}}\), and the viscous force per length, \(f_V\)$\sim$\(\mu U\), effectively vanish because the tangential speed of circular motion $U$ vanishes, which has been justified in the particle tracking experiments of Fig. 4\textit{B}. Therefore, the inertia (other than centrifugal acceleration) cannot be neglected and the coils tend to unwind.

Nevertheless, the coiling radius is constant as shown in Fig. 2\textit{H}. To explain, there are three timescales of interest. Because unwinding typically starts from the free end of the last loop, as shown in Fig. 4\textit{G}, we consider the timescale of unwinding per loop\cite{powers2010dynamics}:
\begin{equation}
\tau_b\sim\frac{\mu}{E}\left(\frac{2\pi R_{\mathrm{coil}}}{d/2}\right)^4
\end{equation}
The time required to wind up a new loop of a coil is $T=\frac{2\pi}{\Omega}\approx\frac{2\pi R_{\mathrm{coil}}}{v_{\mathrm{in}}}$, as shown in the inset of Fig. 4\textit{D}. The timescale of the downward translation of a loop is $\tau_v\sim\frac{R_{\mathrm{coil}}}{v}$. Because $\tau_b\ll\ T$ and $\tau_b\ll\tau_v$, the unwinding process is relatively very slow, and the coiling radius is effectively constant in the timescale of the experiments. To verify our analysis, calculations show that $\frac{\tau_b}{T}\sim$10, which means that 10 new loops should have formed by the time the last loop completes its unwinding. This is consistent with our experimental observations, as shown in Fig. 4\textit{G}.

\section*{Discussion}

This coiling method has several advantages over the conventional pulling methods for analyzing elastic properties of fibers. First, the throughput is thousands of times higher than that of a tensile tester. The bottleneck of the pulling method is, especially for tiny fragile samples, the time-consuming and skill-intensive process of sample loading and unloading. For example, the throughput of tensile testing in our lab is 4 fibers per hour (i.e. 15 min per fiber). Due to the high time and labor cost, the quality of a given batch of fibers can only be assessed through statistical sampling. In comparison, the microfluidic device we are presenting exhibits a significantly higher throughput of at least 3300 fibers per hour, which is 1000 times faster. It is possible to measure the elastic modulus of every microfiber by automated video analysis.

Second, the coiling method enables us to achieve an in situ inline measurement in a microfluidic production line, as demonstrated in Fig. 1, which couples the “making of microfibers” and the “measurement of elastic modulus” on the same process line. Immediately after production, the elastic modulus of every fiber can be measured. Thus, we can screen defective fibers, and use a feedback mechanism to correct the production fault. For example, to maintain a constant modulus, a decrease in coiling radius downstream would trigger an increase in UV intensity upstream.

Third, this method is non-destructive. Unlike in tensile tests, the two end of a sample do not need to be bonded to holders or similar objects. For a fiber with non-uniform elastic modulus, which has been demonstrated in Fig. 3\textit{C}, the conventional method requires cutting the fiber into many shorter segments and measuring each separately. Moreover, we have already justified that the coiling is temporary and does not involve plastic deformation, as shown in Fig. 1\textit{H} and SI Appendix.

We anticipate that the coiling method can be adapted for smaller fibers with submicron diameters by scaling down the setup, adjusting the injection speed and the fluid viscosity. Despite the technical challenges in miniaturizing the microfluidic setup, the task is feasible. There are several successful examples of manipulating and imaging very small fibers, such as DNA and actin filaments, by using micro/nanofluidic chips and optical microscope \cite{chakrabarti2020flexible,chan2004dna,jo2007single,lam2012genome}.

To conclude, the design of our microfluidic device facilitates continuous measurements of the elastic modulus of individual microfibers. By saving the time and labour spent on loading and unloading these fragile samples, the throughput of measurement is a thousand times faster than using a typical tensile tester. The measurement is based on the quantitative relations between the elastic modulus and the buckling length or the coiling radius. The use of either the buckling length or the coiling radius comes with its own set of advantages and disadvantages. For the buckling length method, the result is satisfactory when the fiber has a uniform elastic modulus, but only the modulus of the initial segment is measured. For the coiling radius method, the elastic modulus of the entire fiber is measured. However, we find that a semi-empirical equation constructed with an extra phenomenological parameter gives the best results. Given that the elastic force acting on the coils is much larger than the viscous force and centrifugal force, the coils appear to be stable because the timescale of unwinding is relatively long. We also find that it is inappropriate to deduce the modulus from the pitch because of the lack of a one-to-one correlation. The equation between coiling frequency and coiling radius in micro-channels is the same as that of typical coiling on solid substrate, given that the injection velocity is much faster than the translational velocity of the coils.

Our method can achieve inline measurement in a microfluidic production line, providing real time feedback, enhancing product quality, and reducing costs. We envisage that high-throughput measurements may facilitate potential applications such as screening or sorting by mechanical properties and real-time control during production of microfibers.

\matmethods{\subsection*{Fabrication of fibers by a microfluidic co-flow device}

To fabricate the microfibers, we first prepare a solution that is a ternary mixture of water, polyethylene glycol diacrylate (PEGDA), and tripotassium phosphate \cite{shen2023fibro,lo2019diffusion,vuksanovic2018poly}. It forms an aqueous two-phase system; one polymer-rich phase and one salt-rich phase. After coming to equilibrium, the phases are separated for injection into the microfluidic device. We add 1 wt\% of the photoinitiator, lithium phenyl-2,4,6-trimethylbenzoylphosphinate (LAP), into the polymer-rich phase, which can be solidified by curing with ultraviolet (UV) light.

The device of making fibers is a microfluidic co-flow device with pulsed UV illumination. The co-flow device consists of two coaxially aligned inlets made by a tapered inner capillary in an outer capillary, as shown in Fig. 1\textit{C}, \textit{D}. The polymer-rich phase, which contains the photoinitiator, is injected into the inner capillary, while the salt-rich phase is injected into the outer capillary. The pulsed UV illumination polymerizes the aqueous jet into microfibers as shown in Fig. 1\textit{E}, where we note that the microfiber is straight before entering the wide channel. By adjusting the intensities of UV light, microfibers with different elastic moduli are produced. In the given experiment, the maximum intensity of UV light is 400 mW/cm\textsuperscript{2} and the pulse width is 1 s.

\subsection*{Elastic modulus measurements by a tensile tester}

The elastic moduli of microfibers are measured by a tensile tester (Agilent Technologies T150). During the measurements, the microfibers are immersed in water with 1 wt\% of tripotassium phosphate salt to mimic the environmental condition of the coiling experiments. The elastic moduli are obtained from the slopes of the measured stress-strain curves (see SI Appendix Fig. S4). The strain rate is $10^{-3}$ s\textsuperscript{-1}. For each experimental condition, at least 4 microfibers are measured, and each measurement is repeated 3 times.

\subsection*{Flow visualization by light scattering}
The flow visualization is conducted by adding 0.01wt\% of 10 \textmu m polystyrene particles into the surrounding liquid phase. The particles are illuminated at a large angle, ensuring that only the scattered light enters the camera. The streamlines in Fig. 2\textit{A} are revealed by stacking images of these particles over a period of time.}

\showmatmethods{} 

\acknow{This project is supported by the Research Grants Council of Hong Kong through the Research Impact Fund (No. R7072-18), with H.C.S. as the Project Coordinator, and J.K.N. and H.A.S. as co-Principal Investigators through a research collaboration agreement between the University of Hong Kong and Princeton University. We gratefully thank Alfonso H.W. Ngan and Lei Xu for their help with the experimental instruments. Y.L. and H.C.S. are funded in part by the Health@InnoHK programme from the Innovation and Technology Commission of the Hong Kong SAR government. H.C.S. is funded in part by the Croucher Senior Research Fellowship from Croucher Foundation.}

\showacknow{} 

\bibliography{pnas-sample01}
\newpage

\begin{figure*}
\centering
\includegraphics[width=170mm]{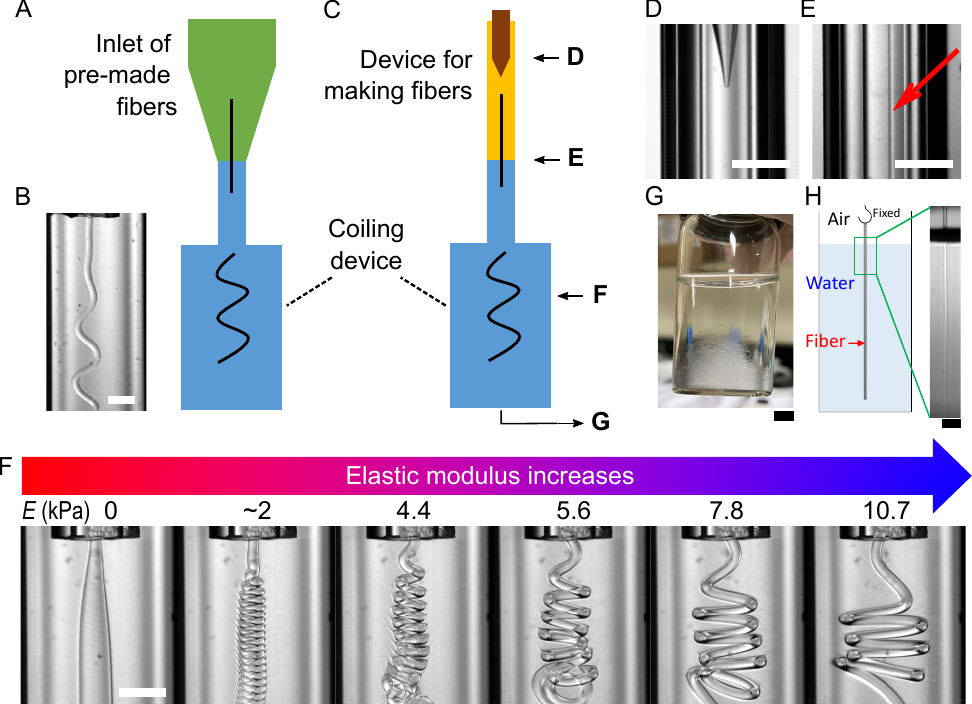}
\caption{\label{figM1}\textbf{Experimental setup and coiling of microfibers.}
(\textit{A}) Drawing of the first setup, which consists of a cone-shaped inlet (green) and a coiling device (blue). The coiling device is made by connecting two glass capillaries to form a small channel upstream (560 \textmu m wide) and an abruptly enlarged wide channel downstream (1746 \textmu m wide). The small channel is a circular tube, while the wide channel is a square tube to minimize image distortion. The setup and the flow are approximately axisymmetric (see SI Appendix); the resulting coiling is three-dimensional.
(\textit{B}) Image of a coiling microfiber (diameter = 120 \textmu m) in the wide channel.
(\textit{C}) Drawing of a second setup in which the coiling device is attached directly to the device for making fibers, which can produce microfibers with different diameters, elastic moduli, and injection speeds on the fly. See also Movie S1 and S2, where microfibers continuously enter the wide channel at a high throughput $v_{\mathrm{in}}$=42 mm/s. The device can run continuously for at least 90 min.
(\textit{D}) Image of the co-flow device, which consists of a tapered inner capillary in an outer capillary.
(\textit{E}) The microfiber (indicated by the arrow) is straight before entering the wide channel downstream.
(\textit{F}) Coiling of microfibers with different elastic moduli, $E$, in the wide channel. No coiling is observed for the liquid jet (first column, $E$=0). The elastic moduli are measured by a tensile tester (see Materials and Methods). See also Movie S3.
(\textit{G}) Collected microfibers in water; $\sim$5000 microfibers are in the container.
(\textit{H}) A collected microfiber is suspended vertically in water, demonstrating that its natural curvature remains zero after temporary coiling.
The scale bar is 5 mm in \textit{G} and 1 mm in \textit{H}. All other scale bars are 500 \textmu m.
}
\end{figure*}
\newpage

\begin{figure*}
\centering
\includegraphics[width=170mm]{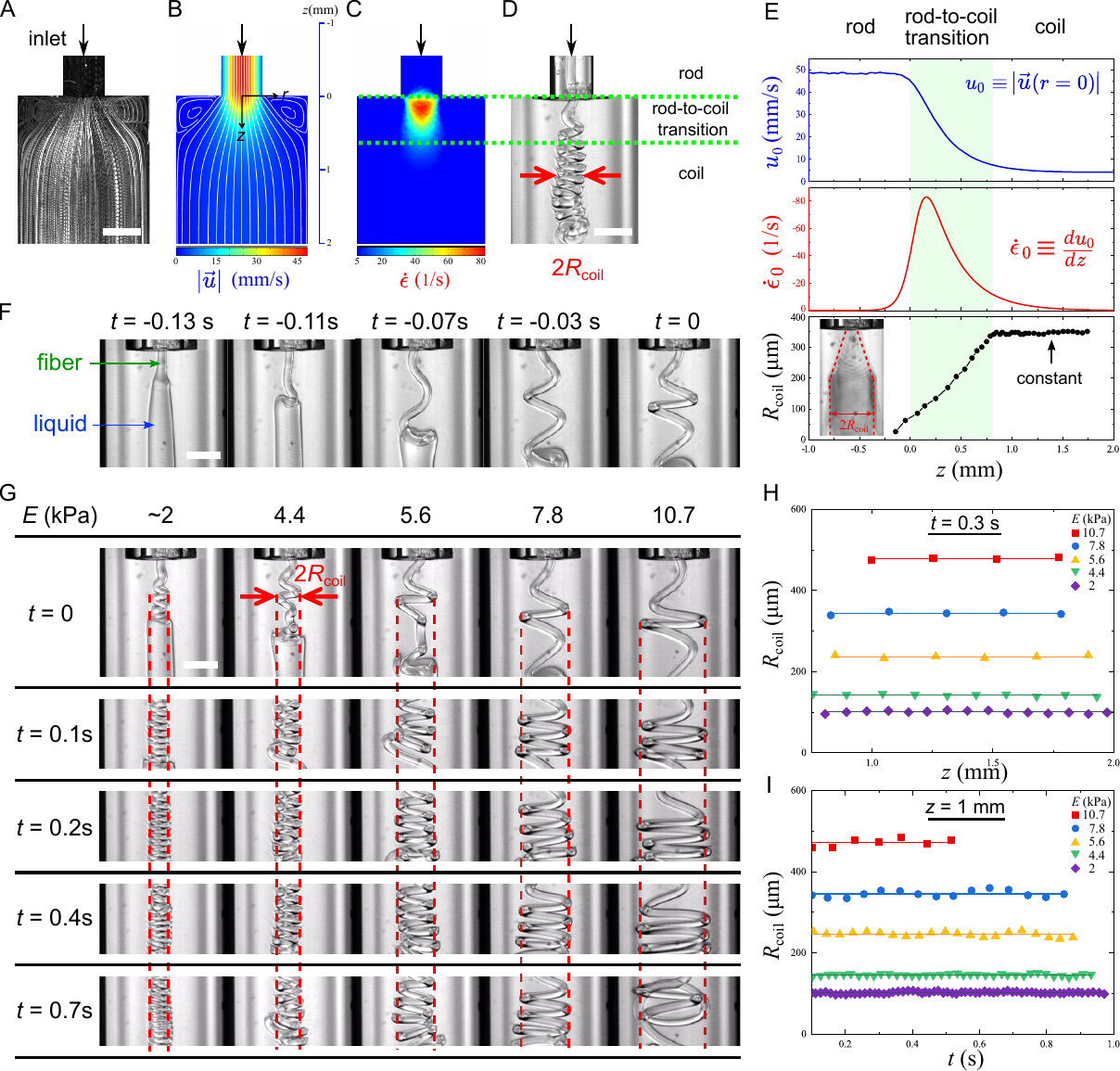}
\caption{\label{figM2}\textbf{General characteristics of system.}
(\textit{A}) Flow visualization of the wide channel.
(\textit{B}) The numerically calculated flow velocity magnitude $\left|\vec{u}\right|$ and streamlines in the absence of fibers. More information about the flow is given in the SI Appendix.
(\textit{C}) The calculated axial strain rate $\dot{\epsilon}=d\vec{u}/dz\cdot\hat{z}$.
(\textit{D}) A photo of a coiling microfiber showing that the rod-to-coil transition occurs in the compressive region.
(\textit{E}) The flow velocity and strain rate along the central line of (\textit{B}) and (\textit{C}). A timelapse photo of the coils is shown in the inset, where the silhouette of the coils (red dotted lines) marks the evolution of coiling radius $R_{\mathrm{coil}}$. While the strain rate increases and then decreases, the coiling radius increases steadily throughout the compression phase. When the localized compression vanishes, the coiling radius becomes constant.
(\textit{F}) Snapshots of the rod-to-coil transition, where the fiber first buckles and then winds into the first coil. We define $t$=0 as the moment where the first coil is formed. The subsequent motion beyond $t$=0 is periodic.
(\textit{G}) Coils of microfibers formed at different times (rows) with different elastic moduli (columns). The elastic moduli $E$ are measured by a tensile tester. All other conditions are the same. The coiling radii $R_{\mathrm{coil}}$ are marked by the red dotted lines.
(\textit{H}) Coiling radii $R_{\mathrm{coil}}$ versus axial position $z$ measured at time $t$=0.3 s. The coiling radii remain constant.
(\textit{I}) Coiling radii $R_{\mathrm{coil}}$ versus time measured at $z$=1 mm. The coiling radii remain constant.
All scale bars are 500 \textmu m.
}
\end{figure*}
\newpage

\begin{figure*}
\centering
\includegraphics[width=170mm]{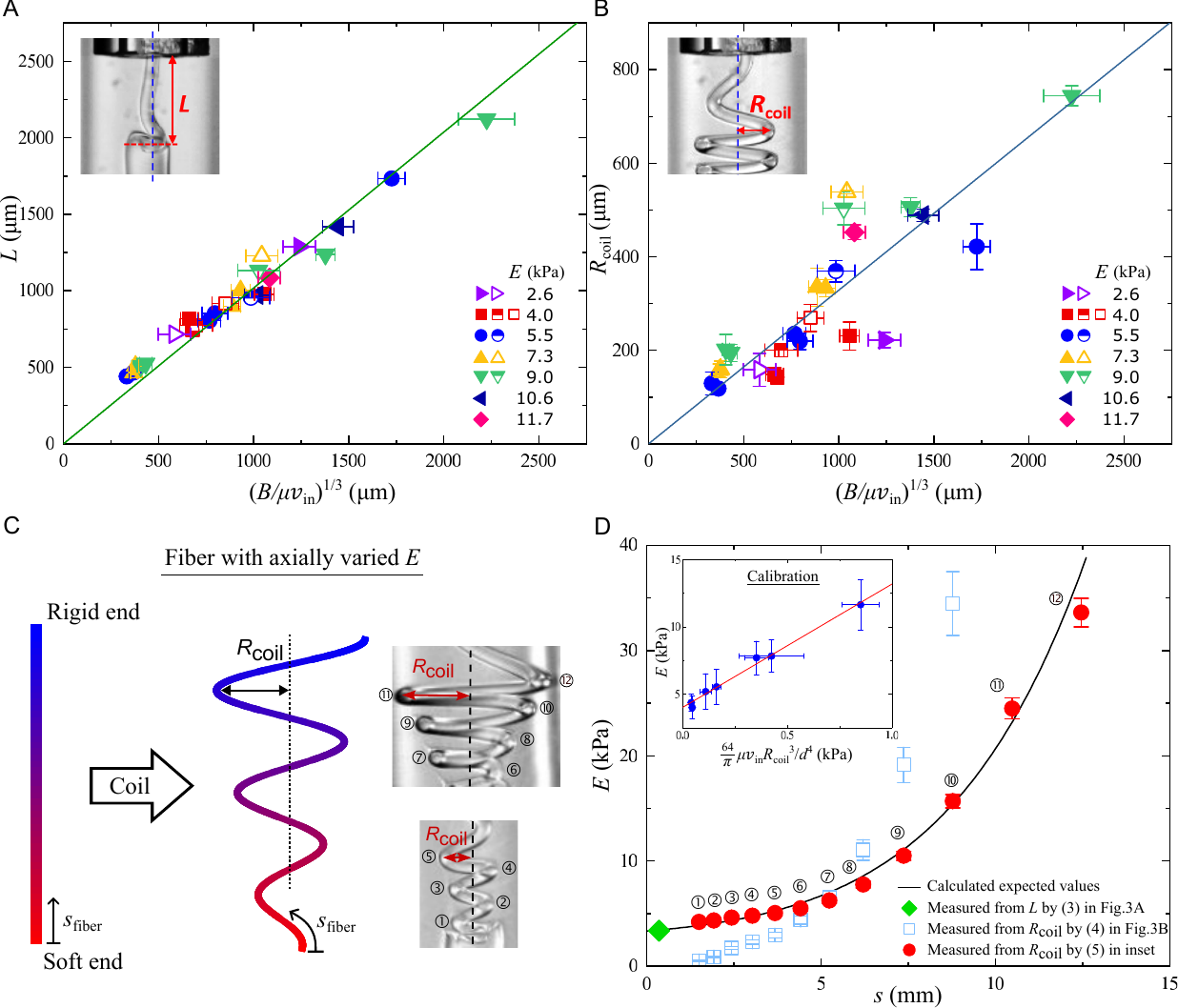}
\caption{\label{figM3}\textbf{Measurement of elastic modulus.}
(\textit{A,B}) Plots of buckling length $L$ and coiling radius $R_{\mathrm{coil}}$ versus $(B/\mu v_{\mathrm{in}})^{1/3}$ for different elastic moduli $E$, injection velocities, and diameters. The solid, half-filled and open symbols represent injection velocities 42, 27 and 15 mm/s respectively. The slopes of the best fit (solid lines) in (\textit{A}) and (\textit{B}) are 1.02 $\pm$ 0.02 and 0.33 $\pm$ 0.02 respectively. The insets illustrate the definition of $L$ and $R_{\mathrm{coil}}$.
(\textit{C}) Illustration of a non-uniform fiber with axially varied elastic modulus. As the fiber coils in the microfluidic device, it forms a sprial with a varying coiling radius that reflects the local elastic modulus. The higher the local elastic modulus, the larger the coiling radius of that segment. See also Movie S4.
(\textit{D}) Comparing the calculated elastic modulus $E$ with that measured from the buckling length $L$ and coiling radius $R_{\mathrm{coil}}$ by equations (3-5). The x-axis $s$ is the arclength from the soft end as defined in \textit{C}. The data points labelled with the number $n$ represent measurements taken from the $n$-th coil labelled in \textit{C}. The inset shows the calibration data and the best fit for (5).
Error bars represent the standard deviation of data.
}
\end{figure*}
\newpage

\begin{figure*}
\centering
\includegraphics[width=170mm]{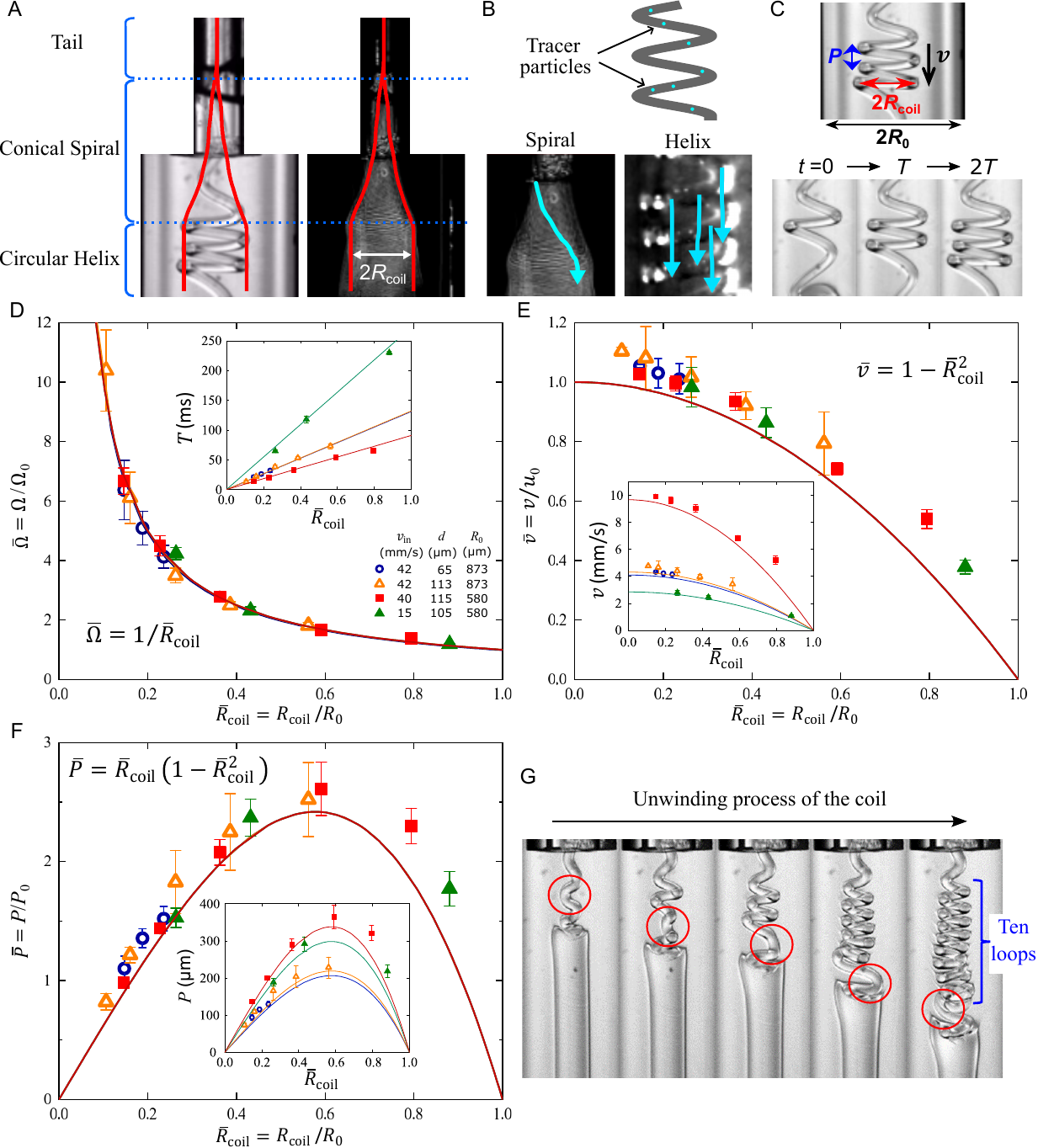}
\caption{\label{figM4}\textbf{Geometric parameters.}
(\textit{A}) Snapshot and time-lapse image, showing the straight tail, the conical spiral and the circular helix. The silhouette is indicated by the red lines.
(\textit{B}) The motion of the spiral and helix are revealed by tracer particles embedded in fibers. For the spiral, the tracers draw a spiral path along the surface of an imaginary cone. For the helix, the tracers are simply translating downwards, leaving straight vertical trajectories.
(\textit{C}) A coil can be fully described by its coiling radius $R_{\mathrm{coil}}$, the pitch $P$, the frequency $\Omega$ (or period $T$), and the downward translational speed $v$. The width of the downstream channel is 2$R_0$. The number of coils increases from one ($t$=0) to two ($t=T$) and three ($t=2T$).
(\textit{D, E, F}) Plots of the normalized coiling frequency $\bar{\Omega}$, translational velocity $\bar{v}$ and pitch $\bar{P}$ versus the normalized coiling radius ${\bar{R}}_{\mathrm{coil}}$ under different experimental conditions. The solid lines represent the equations written next to them. The insets show the data without normalization. The scaling terms are $\Omega_0 \equiv v_{\mathrm{in}}/{R_0}$, $u_0\equiv|\vec{u}\left(r=0\right)|$ and $P_0\equiv2\pi u_0R_0/v_{\mathrm{in}}$.
(\textit{G}) A photo series that captures the slow unwinding process of the coil.
}
\end{figure*}

\end{document}